# THz-circuits driven by photo-thermoelectric graphene-junctions

**Andreas Brenneis[1,2], Felix Schade[1,2], Simon Drieschner[1,2], Florian Heimbach[2,3],**

**Helmut Karl[4], Jose A. Garrido[1,2], and Alexander W. Holleitner[1,2]***

[1] Walter Schottky Institut and Physics Department, Technical University Munich, Am Coulombwall 4a, 85748 Garching, Germany.

[2] Nanosystems Initiative Munich (NIM), Schellingstr. 4, 80799 Munich, Germany.

[3] Lehrstuhl für Physik funktionaler Schichtsysteme, Physics Department, Technical University of Munich, D-85748 Garching, Germany

[4] Institute of Physics, University of Augsburg, 86135 Augsburg, Germany.

*corresponding author: holleitner@wsi.tum.de

## ABSTRACT

**For future on-chip communication schemes, it is essential to integrate nanoscale materials with an ultrafast optoelectronic functionality into high-frequency circuits. The atomically thin graphene has been widely demonstrated to be suitable for photovoltaic and optoelectronic devices because of its broadband optical absorption and its high electron mobility. Moreover, the ultrafast relaxation of photogenerated charge carriers has been verified in graphene. Here, we show that dual-gated graphene junctions can be functional parts of THz-circuits. As the underlying optoelectronic process, we exploit ultrafast photo-thermoelectric currents. We describe an immediate photo-thermoelectric current of the unbiased device following a femtosecond laser excitation. For a picosecond time-scale after the optical excitation, an additional photo-thermoelectric contribution shows up, which exhibits the fingerprint of a spatially inverted temperature profile. The latter can be understood by the different time-constants and thermal coupling mechanisms of the electron and phonon baths within graphene to the substrate and the metal contacts. The interplay of the processes gives rise to ultrafast electromagnetic transients in high-frequency circuits, and it is equally important for a fundamental understanding of graphene-based ultrafast photodetectors and switches.**

Graphene is a gapless broadband absorber, which allows to detect wavelength-multiplexed signals in the frequency range from THz to UV-Vis-radiation.[1] In particular, free-charge carrier excitations can be exploited for THz signal propagation[2], whereas interband excitations, like photogenerated electron-hole pairs, can drive ultrafast currents because of a femtosecond relaxation of photogenerated charge-carriers. A corresponding ultrafast photovoltaic current is driven by built-in electric fields, for instance, at metal-graphene junctions[3,4,5,6,7] and by an externally applied bias voltage.[8] An alternative route to harvest and exploit the energy of incident photons are so-called photo-thermoelectric currents, which have been reported for graphene *p-n* junctions[9,10,11,12,13] and junctions made from mono- and bilayer graphene.[14] The thermopower differs on the two adjacent sides of these junctions, and the optical excitation generates the necessary heat to drive the thermoelectric current. Graphene is particularly suited to host thermoelectric currents, because the electron-electron scattering is the dominant ultrafast decay channel for a low irradiation intensity.[15] In turn, a hot electron distribution can quickly form after the optical excitation with only a small dissipation to the crystal lattice.[13,16,17,18] Hereby, non-equilibrium charge carrier temperatures of $10^3$-$10^4$ K can be achieved.[15,19] Such an elevated temperature in combination with the large thermopower of graphene (≈100 µV/ K) gives rise to a significant photo-thermoelectric response. As it is shown here, the latter can be used for triggering ultrafast signals in high-frequency circuits.

We exploit an on-chip pump-probe photocurrent spectroscopy based on THz-waveguides and Auston-switches with a sub-picosecond time-resolution.[7,20,21,22] Our technique can detect photocurrents that are linear with respect to the laser power,[13,15] and moreover, we detect signals which trigger electromagnetic transients in the THz circuits.[20,23,24] We can distinguish two ultrafast contributions in the graphene-junctions. On the one hand, an immediate response occurs during the pulsed excitation. This immediate response exhibits a six-fold sign-change as a function of both gate voltages of the dual-gated junction. This is a clear finger-print of a photo-thermoelectric current.[9,11,18] On the other hand, we observe a decaying photocurrent contribution on the time-scale of a few picoseconds. We can numerically compute this decaying response by a thermoelectric current which is driven by an inverted temperature profile within the graphene device. The non-equilibrium temperature inversion occurs because the electron and phonon baths of the graphene couple differently to the metal contacts and to the underlying substrate of the device. Our results show that the heat loss to the immediate environment of the graphene plays an important role in optimizing the energy conversion efficiency in graphene. The insights into the non-equilibrium carrier dynamics may proof essential for future ultrafast optoelectronic switches and for the fundamental understanding of photovoltaic devices and photodetectors. In particular, our work experimentally proves that photo-thermoelectric currents can be used for an ultrafast optoelectronic signal-conversion in graphene as a functional part of high-frequency circuits.[25,26]

## Results

**Sample Fabrication**

Figure 1a sketches the sample layout of the dual-gated graphene junction. The circuit provides two gates labeled gate1 and gate2. Applying corresponding gate voltages $V_{g1}$ and $V_{g2}$ controls the Fermi levels and thus the thermopower in the graphene on each side of the junction. The graphene is electrically contacted by metal strips made from Ti/Au (10 nm/250 nm), which allow to apply a bias voltage $V_{sd}$. First, two adjacent 10 nm thick Pt gates are lithographically defined on a sapphire substrate. Then, a 30 nm thick, ALD-grown (atomic layer deposition) $Al_2O_3$-layer isolates the gates from the graphene. Subsequently, CVD-grown (chemical vapor deposition) graphene is mechanically transferred to the sample.[27] The graphene region of interest is 25 µm wide and meanwhile protected with a resist, while the remaining graphene is removed by utilizing an oxygen plasma. In a final lithography step, the protective resist is removed by a solvent, and the striplines are lithographically defined on top of the graphene. Figure 1b shows an optical microscope image of the device with the graphene (striplines) being highlighted in blue (yellow). The gap between the two underlying Pt-gates is clearly visible as dark contrast. Moreover, the Ti/Au-contacts form coplanar high-frequency striplines with a total length of ≈44 mm, a width of 5 µm, and a separation of 10 µm. The striplines are needed to measure the time-resolved photocurrent $I_{sampling}$.[7,20]

**Scanning photocurrent microscopy**

For the excitation of the graphene, we use a laser system that emits laser pulses with a 30 fs pulse duration at a repetition rate of 80 MHz in the spectral range from 900 to 1500 nm (cf. Supplementary Information). The laser beam is focused with a 15x reflective objective. The laser spot on the surface of the graphene has a full width half maximum (FWHM) of ≈ 5.5 µm. All presented experiments are performed with an excitation power of 1-2 mW at room temperature and in a vacuum chamber. First, the time-integrated photocurrent $I_{photo}$ is measured with the help of a current-voltage amplifier while the sample is scanned laterally by the help of piezoelectric positioners. Figure 1c depicts such a photocurrent map of $I_{photo}$ at $V_{sd}$ = $V_{g1}$ = $V_{g2}$ = 0V. For these experimental parameters, $I_{photo}$ shows opposite signs on both sides of the junction, as it will be discussed below.

**THz-time domain photocurrent spectroscopy**

For measuring the ultrafast optoelectronic dynamics of the graphene-based junction, we use an on-chip THz-time domain photocurrent spectroscopy of the stripline circuits (Figure 1d),[7,20,21,22] where an optical femtosecond pump laser excites the charge carriers in the graphene (Figure 1e). This laser is the same as for the time-integrated photocurrent measurements (Figure 1a). Since the contacts form striplines, the photocurrents give also rise to electromagnetic transients in the metal striplines with a bandwidth of up to 2 THz.[20,21] The transients run along the striplines, and they are detected on-chip by a time-delayed optical femtosecond probe pulse in combination with an Auston-switch (Figure 1e).[20] We use ion-implanted amorphous silicon for this ultrafast photodetector with a time-resolution of about 1 ps.[21,22] The probe pulse has a photon energy of 1.59 eV and a temporal width of 100 fs. The current $I_{sampling}$ across the Auston-switch samples the electromagnetic transients on the striplines as a function of the time-delay $\Delta t$ between the two laser pulses, and it is directly proportional to the ultrafast photocurrents in the graphene. Hereby, the photocurrents in the graphene can be measured with a picosecond time-resolution.[7,20]

**Resistance properties of the graphene junction**

The two-terminal resistance $R_{\text{total}}$ is measured with a current-voltage amplifier to gain insights into the electronic properties of the dual-gated graphene junction. Figure 2a shows the experimental dependence of $R_{\text{total}}$, when both gate1 and gate2 are varied simultaneously. We find the charge neutrality point of the graphene for negative gate voltages, which implies that it is intrinsically $n$-doped. Figure 2b shows $R_{\text{total}}$, when $V_{g1}$ and $V_{g2}$ are varied independently. We make use of $\sigma = \frac{l}{wR_{\text{total}}}$ to express $\sigma$ as a function of the measured total resistance $R_{\text{total}}$, with $l = 10$ µm and $w = 25$ µm the length and width of the two-terminal graphene circuit. We model the charge carrier density as

$$n = \sqrt{n_{\text{ind}}^2(V_{gi}) + n_0^2} = n_0\sqrt{\left(\frac{V_{gi}-V_{sd}}{\Delta}\right)^2 + 1}, \qquad (1)$$

to obtain $R_{\text{total}} = \frac{l}{w}\frac{1}{ne\mu}$ for each gate voltage $V_{gi}$ (with $i = 1, 2$).[29] Here, $n_0$ is the residual charge carrier density, $n_{\text{ind}}$ the gate-induced charge carrier density for gate1 and gate2, $n$ the total charge carrier density, and $\mu$ the mobility. The term $\Delta$ is defined as $\Delta = \frac{n_0 e}{C^*}$. $C^*$ describes the specific capacitance $C^* = \frac{\varepsilon_0 \varepsilon_r}{d}$ between the two gates and the graphene-sheet[29] (with $\varepsilon_r = 6.0$ the relative permittivity and $d = 30$ nm the thickness of the Al$_2$O$_3$ layer). The data in Figure 2b can be fitted by the following expression

$$R_{\text{total}}(V_{g1}, V_{g2}) = R_{\text{contact}} + \frac{1}{w}\sum_{i=1,2}\frac{l_i}{\mu_i \Delta_i C^*}\sqrt{1 + \left(\frac{V_{gi}-V_{di}}{\Delta_i}\right)^2}^{-1}, \qquad (2)$$

with $l_{1,2} = 4$ µm the length of both gated areas. The expression $R_{\text{contact}}$ accounts for the constant resistance of the metal striplines and the ungated graphene region. Our fit gives $R_{\text{contact}} = 2.57$ kΩ and mobilities of $\mu_1 = 157$ cm$^2$ V$^{-1}$s$^{-1}$ and $\mu_2 = 278$ cm$^2$ V$^{-1}$s$^{-1}$. The charge neutrality point is reached at $V_{d1} = -1.92$ V and $V_{d2} = -2.77$ V. The fit further reveals $\Delta_1 = 3.20$ V and $\Delta_2 = 2.62$ V. Figure 2c shows the resulting fit (cf. Supplementary information).

**Photo-thermoelectric properties**

The photo-thermoelectric effect[9,10,11,18] and the photovoltaic effect[3,4,6,7,8] are considered as the two distinctive photocurrent generation mechanisms in graphene on short time-scales. The photovoltaic effect relies on a drift current, where photogenerated charge carriers are separated by an electric field. In turn, the photovoltaic current monotonically depends on the difference of the Fermi levels of both sides of a dual-gated graphene junction (cf. Supplementary Information). In the case of a thermoelectric generation mechanism, the photocurrent is a non-monotonic function. In particular, the thermoelectric voltage $V_{\text{TE}}$ can be written as

$$V_{\text{TE}} = \int \frac{\partial T}{\partial x} \cdot S(x) \cdot dx = \sum_i \Delta T_i \cdot S_i(V_{gi}) = \Delta T \cdot [S_{g1}(V_{g1}) - S_{g2}(V_{g2})], \qquad (3)$$

with $\Delta T_i$ the change in electron temperature for a segment with constant thermopower $S_i$. We define the x-axis as in Figure 1c. The expression can be simplified to the last term of eq. (3) for a symmetric heat profile with amplitude $\Delta T$ and for constant thermopowers $S_{g1}$ and $S_{g2}$ for the two areas on top of gate1 and gate2. We note that a symmetric heat profile is introduced by the laser pulse with the given FWHM of ≈ 5.5 µm (Figure 1a). The thermopower or so-called Seebeck coefficient can be related to the electric properties of the device following the Mott formula[28]

$$S = -\frac{\pi k_B^2 T_0}{3e} \cdot \frac{1}{\sigma} \frac{\partial \sigma}{\partial E}\bigg|_{E=E_F} = \frac{\pi k_B^2 T_0}{3e} \cdot \frac{1}{R} \cdot \frac{\partial R_{\text{total}}}{\partial E}\bigg|_{E=E_F}, \qquad (4)$$

with $E_F$ the Fermi-energy, $k_B$ the Boltzmann constant, $e$ the elementary charge, $\sigma$ the electric conductivity, and $T_0 = 300$ K the equilibrium electron temperature. With the definitions from equation (1) and $E_F = \pm \hbar v_F \sqrt{\pi |n_{\text{ind}}|}$, eq. (4) can be rewritten for each gate voltage

$$S_{gi} = \frac{2k_B^2 T_0}{3R_{\text{total}} \hbar v_F} \cdot \sqrt{\frac{\pi}{eC^*} \cdot |V_{gi} - V_{sd}|} \cdot \frac{\partial R_{\text{total}}}{\partial V_{gi}} \qquad (i=1,2). \qquad (5)$$

Eq. (5) is used to calculate the thermopower (Seebeck coefficients) $S_{gi} = S_{gi}(V_{gi})$ from the fit of $R_{\text{total}} = R_{\text{total}}(V_{g1}, V_{g2})$ in Figure 2c. Figure 2d allows to intuitively understand the non-monotonic behavior of the thermopower as a function of $V_{g1}$ and $V_{g2}$: for a given gate voltage $V_{g2}$ and Seebeck coefficients $S_{g2}$ (black arrow in Figure 2d), there are two possible gate voltages $V_{g1}$, which result in a coincident Seebeck coefficient $S_{g1} = S_{g2}$. In turn, the thermoelectric current will show two crossing points as a function of $V_{g1}$ and $V_{g2}$ according to the right-hand-side of eq. (3). The resulting six-fold sign change is therefore a characteristic, non-monotonic fingerprint of a thermoelectric current (Figure 2e).[9,11,18] In the calculus, the gate voltages are varied from -14 to +14 V. Experimentally, the breakdown voltage of the $Al_2O_3$-layer limits the maximum sweeping range to ± 7 V (dashed square in Figure 2e). Figure 2f presents the corresponding time-integrated photocurrent $I_{photo}$ of the dual-gated junction as a function of $V_{g1}$ and $V_{g2}$, when the exciting laser is focused onto the center of the graphene junction (circle in Figure 1c). Given the limited gate sweeping range, we clearly observe the sign change as expected from the pattern of the calculated thermoelectric current in Figure 2e.

**Ultrafast currents in the graphene-junction**

In the following, we discuss the temporal dynamics of the non-equilibrium photocurrents in the graphene junction as detected by the THz stripline circuit. Figure 3a depicts the time-resolved photocurrent $I_{sampling}(\Delta t)$ for different gate voltages $V_{g1}$ and $V_{g2}$. The voltage settings are defined by the indices in Figure 3b. The signal can be described by a 'peak current' as an immediate response (green area in Figure 3a), that occurs when the delay $\Delta t$ between pump and probe pulse is zero. For a longer time-delay, we observe a rising and decaying behavior of $I_{sampling}$ (gray area). We label the latter response as 'decay current'. For each gate voltage setting, we fit $I_{sampling}(\Delta t)$ with a constant offset, a Lorentzian (green area), and an exponentially rising and decaying function $\propto \left( \exp\left[\frac{-\Delta t}{\tau_{decay}}\right] - \exp\left[\frac{-\Delta t}{\tau_{rise}}\right] \right)$ (gray area). The latter is a phenomenological function to describe the form of the decay current with rise and decay times $\tau_{decay}$ and $\tau_{rise}$.[20] They will be discussed below. The temporal width of the Lorentzian is 1.28 ± 0.18 ps, which coincides with the bandwidth limitation of the present stripline circuit.[30] The bandwidth is limited by the dispersion and attenuation of the striplines. The dependences of the peak and decay currents on the gate voltages are depicted in Figure 3b and 3c, respectively. The peak current exhibits a manifold sign change, which is in agreement with a thermoelectric current within the limited gate voltage range (compare to the highlighted square in Figure 2e). The decay current rather depends on $V_{g2}$ than on $V_{g1}$ for this laser position. We further note that the peak current and the decay current can have different polarities at certain gate voltages (e.g. setting 1, 3, and 4 in Figure 3a). At the given signal-to-noise-ratio, however, the sum of the peak and decay current (Figure 3d) mimics the time-integrated signal $I_{photo}$ (Figure 2f) to a better extend than the peak current alone (Figure 3b). This finding is consistent with the fact that the time-integrated signal comprises the sum of all ultrafast currents.

## Discussion

In order to explain the peak and the decay currents, we consider the ultrafast charge carrier relaxation dynamics. After the pulsed excitation, the charge carriers relax due to electron-electron scattering on a femtosecond time-scale, and the carrier ensemble can quickly be described by a hot Fermi-Dirac distribution.[13,15,18,31,32,33,34,35] The electron and phonon bath temperatures equilibrate on a picosecond time-scale[13,16,17,18] due to optical phonon scattering,[36,37] surface polar phonon scattering,[38,39] and supercollision scattering.[40,41,42] From the striking similarity of the expected pattern of the thermoelectric current in Figure 2e and the measured dependence of the 'peak current' in Figure 3b, we can conclude that this immediate photocurrent is dominated by an ultrafast thermoelectric current at the junction. The necessary non-equilibrium electron temperature can be assumed to be as high as $10^4$ K. [15,19] We find that it is important to excite at the center of the dual-gated junction such that a symmetric heat profile is impinged by the Gaussian laser profile with its given FWHM (sketch in Figure 1a and circle in Figure 1c). Hereby, the balanced temperature profile ($|\Delta T_1| = |\Delta T_2| = \Delta T$) and the difference in the thermopowers on both sides of the graphene give rise to a thermoelectric voltage $V_{TE} = (S_{g1} - S_{g2}) \cdot \Delta T$, that drives a photocurrent as long as the temperature profile is present. We can further infer that the so-called photovoltaic effect plays only a negligible role. In particular, the photovoltaic effect would only depend on the difference of both Fermi levels of dual-gated graphene junctions. In other words, its sign should change along the diagonals of Figures 3b and 3c, which we do not observe (cf. Supplementary Information). We can also exclude a photo-Dember current,[43,44] because its amplitude should only depend on the difference of the electron- and hole-mobilities without a sign-change as in Figures 3b and 3c. Instead, we demonstrate below that the observed decay current is of thermoelectric origin as well.

Figures 4a-c show the area of the decay current for a time-delay of 3.3 ps $< \Delta t <$ 13.3 ps as a function of both gate voltages for three different excitation positions of the laser spot. The excitation position is varied in steps of 2 µm from an excitation position closer to gate1 (Figure 4a), to one at the center between both gates (Figure 4b), and to one closer to gate2 (Figure 4c). We numerically reproduce the corresponding patterns by equation eq. (3) (Figures 4d, 4e, 4f), if we consider a slight shift of the charge neutrality point and an inverted temperature profile (black dots in Figures 4g, 4h, and 4i). The shift of the charge neutrality point is induced by the position of the laser spot, because depending on the position, the static, optically induced background doping changes. For instance, the pattern of Figure 4a (4c) with the laser spot on gate1 (gate2) shows a predominant dependence on $V_{g1}$ ($V_{g2}$). In other words, the decay current predominantly depends on the voltage of the gate at which the laser is focused. To explain the inverted temperature profiles, we need to consider the phonon bath of the graphene. It is heated on a picosecond time-scale by the mentioned electron-phonon scattering processes.[16,17,18,36,37,38,39,40,41,42] The heat will then be transferred to the environment of the graphene. In the case of the underlying $Al_2O_3$-layer and sapphire substrate, this will occur on a 1-10 ps time-scale.[22,45,40] On this time-scale, the temperature of the electron and phonon baths adopt the same values. In this sense, we interpret $\tau_{decay}$ and $\tau_{rise}$ of the decay current. For the metal striplines, the heat transfer occurs on a similar time-scale. However, the heat diffusion along the striplines is slower with a time-scale of up to 150-300 ps.[7] In turn, the electron bath in the center of the dual-gated graphene junction efficiently cools-down faster than at the sections which are in vicinity to the metals. In turn, a thermoelectric current is generated with a corresponding longer time-scale at positions closer to one of the contacts, especially if the contacts are heated by the pump pulse.[7,22] We note that both the 'peak current' and the 'decay current' are detected via our THz-time domain photocurrent spectroscopy. Therefore, both trigger ultrafast optoelectronic signals in the high-frequency circuit, and they can be switched by the two gate voltages of the graphene-junction (Figures 3 and 4).

In conclusion, we show that the ultrafast photocurrent in a dual-gated graphene junction is dominated by the photo-thermoelectric effect. The Fermi energy of the two sides of the junctions is controlled by two adjacent back gates. Surprisingly, we find two distinct ultrafast thermoelectric contributions. A first immediate response occurs within the temporal resolution of our experiment, which is about one picosecond. An unambiguous dependence on the two gate voltages identifies this peak current as an ultrafast photo-thermoelectric current. In our interpretation, this current occurs on a sub-ps time-scale after the pulsed femtosecond laser excitation when the electrons form a hot Fermi Dirac distribution due to fast electron-electron scattering. A second ultrafast thermoelectric response is identified with a decaying constant of a couple of picosecond. We interpret the latter time-scale to represent the heat coupling of the electron and phonon baths to the environment of the graphene. We can numerically reproduce the second response by a thermoelectric current as long as the heat baths are out of equilibrium. The sum of the two time-resolved photocurrent contributions resembles the time-integrated photocurrent measured in the dual-gated graphene junction. The findings may proof essential for graphene-based photodetectors and switches integrated in future high-frequency circuits.

## Conflict of Interest

The authors declare no competing financial interest.

## Corresponding Author

holleitner@wsi.tum.de

## Author Contributions

A.B. and A.W.H. designed the experiment. A.B., F.S., A.W. performed the experiments. A.B., F.S., S.D., F.H., H.K., J.A.G., and A.W.H engineered the materials and the stripline geometry. The manuscript was written through contributions of all authors. All authors have given approval to the final version of the manuscript.

## Acknowledgements

We thank D. Turchinovich for fruitful discussions and C. Karnetzky for technical assistance. The work was financially supported by the European Research Council (ERC) Grant NanoREAL (No. 306754).

## Conflict of Interest

The authors declare no competing financial interest.

## Corresponding Author

holleitner@wsi.tum.de

## Author Contributions

A.B. and A.W.H. designed the experiment. A.B., F.S., A.W. performed the experiments. A.B., F.S., S.D., F.H., H.K., J.A.G., and A.W.H engineered the materials and the stripline geometry. The manuscript was written through contributions of all authors. All authors have given approval to the final version of the manuscript.

## Acknowledgements

We thank D. Turchinovich for fruitful discussions and C. Karnetzky for technical assistance. The work was financially supported by the European Research Council (ERC) Grant NanoREAL (No. 306754).

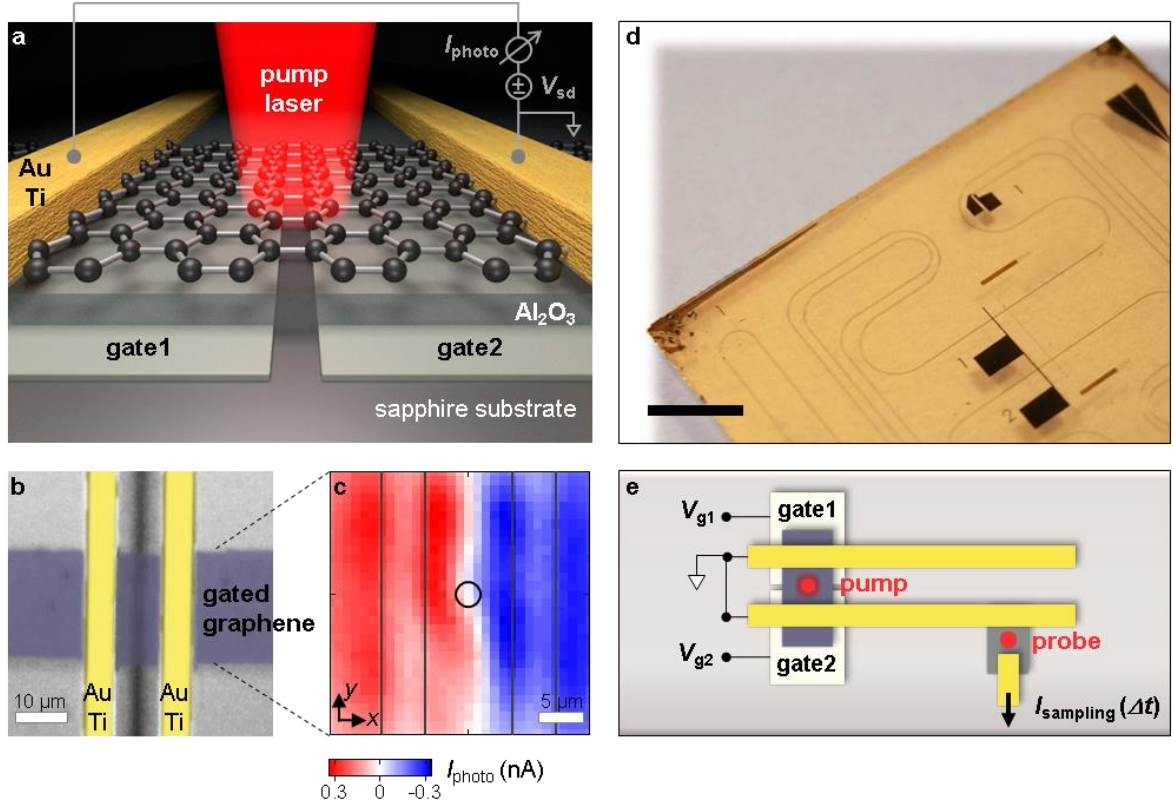

**Figure 1. Gated graphene-junction integrated in a THz circuit.** (a) Sketch of a dual-gated graphene junction with two bottom gates, named gate1 and gate2, on top of a sapphire substrate and an insulating $Al_2O_3$-layer in-between. The graphene is contacted by two Ti/Au-striplines. A pulsed laser generates charge carriers in the graphene. (b) Microscope image of a corresponding graphene sheet (highlighted in blue) and the striplines (yellow) on top of two gates. The lateral gap between both gates is ≈2 µm (dark region in the center). (c) Spatially resolved and time-integrated photocurrent $I_{photo}$ of the graphene sheet shown in Figure 1b as a function of the coordinates $x$ and $y$. For $x = y = 0$ µm, the sample is excited at the positon of the circle. The vertical contour lines depict the positions of the striplines. (d) Image of the overall THz-stripline circuit on a sapphire chip. Scale bar, 2 mm. (e) Sketch of the on-chip pump/probe scheme. A pump pulse excites the gated graphene-junction, while the probe pulse triggers an Auston-switch. The latter is an ultrafast photoswitch and it reads out the electromagnetic transients of the striplines. As a function of the time-delay $\Delta t$, the corresponding current $I_{sampling}$ gives insights into the electromagnetic signals running along the high-frequency striplines and hereby, into the temporal optoelectronic dynamics of the graphene-junction.

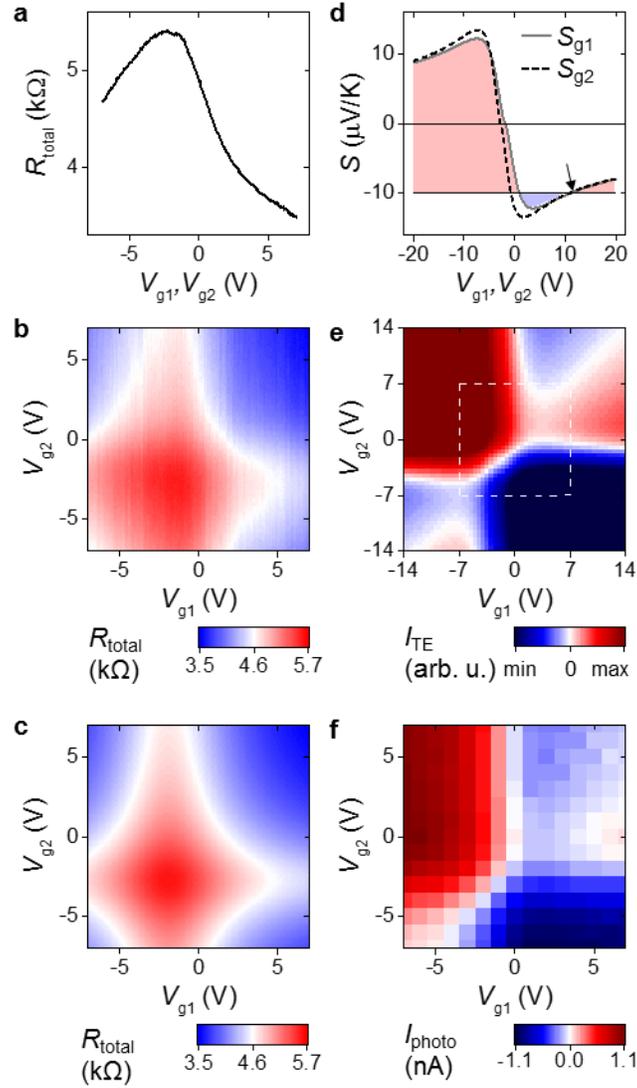

**Figure 2 Electronic and photo-thermoelectric properties of the graphene-junction.** (a) Two-terminal resistance $R_{\text{total}}$ of the dual-gated graphene junction, when the two gate voltages $V_{g1}$ and $V_{g2}$ are swept simultaneously. The maximum gives the charge neutrality point of the graphene sheet. (b) Two-dimensional resistance plot $R_{\text{total}} = R_{\text{total}}(V_{g1}, V_{g2})$, when the gates are varied individually. (c) Two-dimensional fit of $R_{\text{total}}$ according to eq. (2). See text for details. (d) Seebeck coefficient $S_{g1}$ ($S_{g2}$) vs. $V_{g1}$ ($V_{g2}$) calculated from the fit parameters of $R_{\text{total}}$ according to eq. (5). (e) Calculated thermoelectric current vs. $V_{g1}$ ($V_{g2}$) for the doping and mobility parameters from (c) according to equation eq. (3) and $I_{\text{TE}} = V_{\text{TE}}/R_{\text{total}}$. (f) Experimental time-integrated photocurrent $I_{\text{photo}}$ vs. $V_{g1}$ ($V_{g2}$) at $x = y = 0$ (circle in Figure 1c).

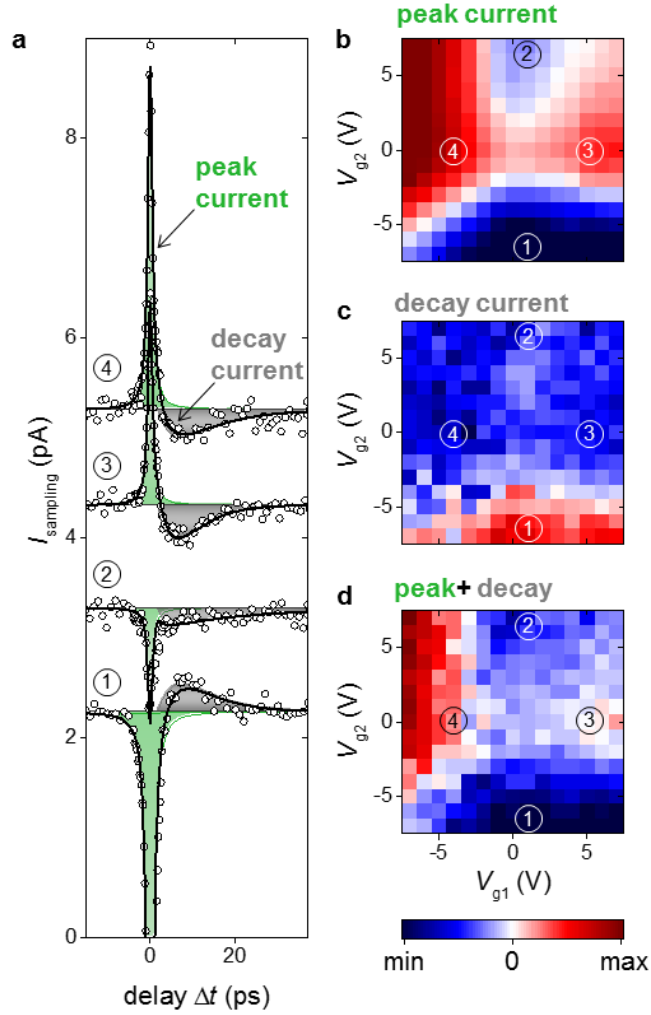

**Figure 3. Ultrafast photo-thermoelectric currents in the graphene-junction measured by the THz-circuit.** (a) Time-resolved photocurrent $I_{\text{sampling}}(\Delta t)$ for four different gate voltages at the position $x = y = 0$ (circle in Figure 1c) with a distinctive 'peak current' (green area) and a 'decay current' (gray area). (b) Area of the peak current vs. $V_{g1}$ and $V_{g2}$ with the four gate voltage settings 1, 2, 3, and 4 of (a). (c) Area of the decay current vs. $V_{g1}$ and $V_{g2}$. (d) Sum of the areas in figures (b) and (c).

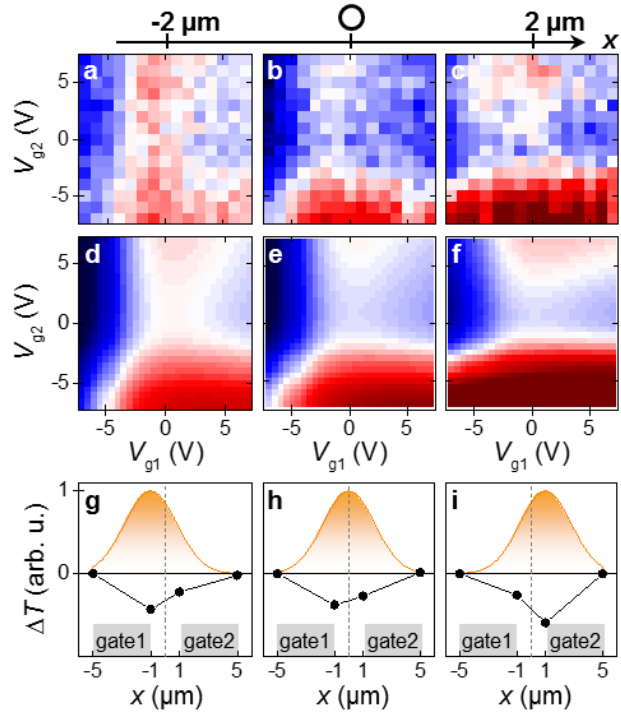

**Figure 4 Heat profiling of the graphene-junction.** (a) to (c): Area of the decay current (gray area in Figure 3a) for three different positions of the laser spot ($x$ = -2, 0, 2 µm, and $y$ = -7 µm). (d) to (f): Thermoelectric current calculated for an inverted temperature profile defined in the figure (g), (h), and (i). (g) to (i): Inverted temperature profile (black line) across the graphene junction for time-scales $\Delta t > 1$ ps. The orange area depicts the Gaussian excitation intensity of the exiting laser pulse, i.e. the initial temperature profile at $\Delta t = 0$ ps.

**Publication bibliography**

# THz-circuits driven by photo-thermoelectric graphene-junctions

**Andreas Brenneis[1,2], Felix Schade[1,2], Simon Drieschner[1,2], Florian Heimbach[2,3], Helmut Karl[4], Jose A. Garrido[1,2], and Alexander W. Holleitner[1,2]***

[1] Walter Schottky Institut and Physics Department, Technical University Munich, Am Coulombwall 4a, 85748 Garching, Germany.
[2] Nanosystems Initiative Munich (NIM), Schellingstr. 4, 80799 Munich, Germany.
[3] Lehrstuhl für Physik funktionaler Schichtsysteme, Physics Department, Technical University of Munich, D-85748 Garching, Germany
[4] Institute of Physics, University of Augsburg, 86135 Augsburg, Germany.
*corresponding author: holleitner@wsi.tum.de

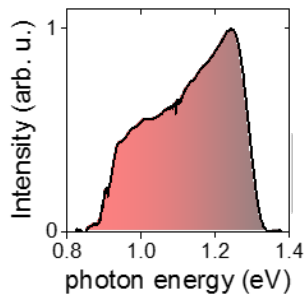

**Figure S1:** Spectra of the pump laser pulse that is used to excite the graphene at the junction between both gates.

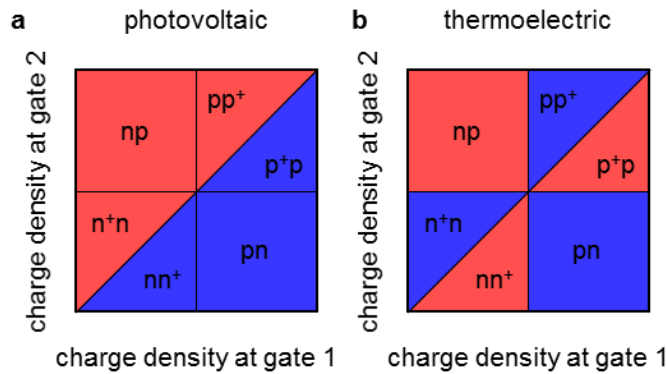

**Figure S2:** Sketch of the dependence of (a) the photovoltaic and (b) the thermoelectric photocurrent on the charge density in two adjacent graphene regions (red: positive current, blue: negative current).[1]

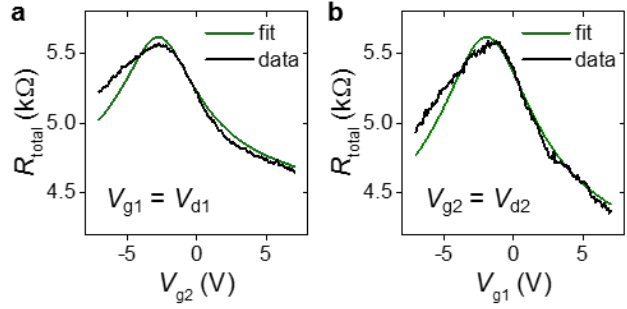

**Figure S3:** Fit of $R_{\text{total}}(V_{g1}, V_{g2})$ for (a) charge neutrality at gate1 ($V_{g1} = V_{d1} = -1.9V$) and (b) charge neutrality at gate2 ($V_{g2} = V_{d2} = -2.8V$).

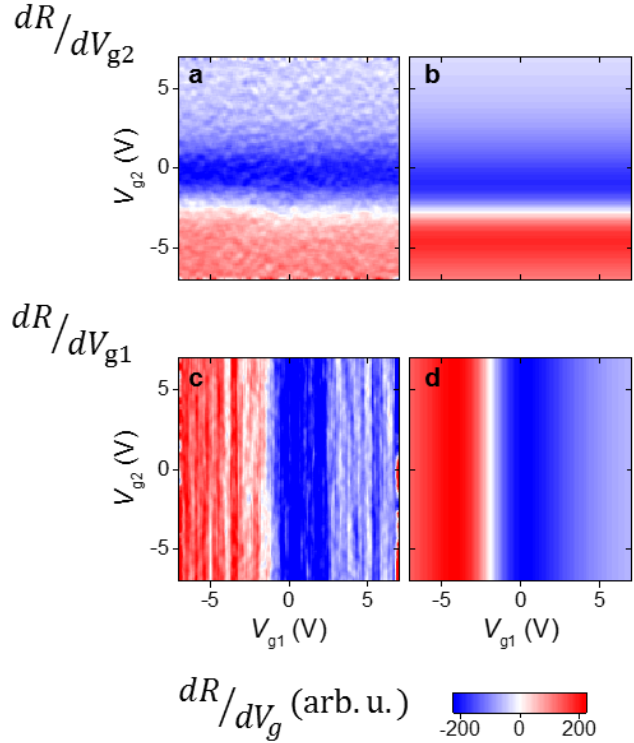

**Figure S4:** Calculation of the thermopower from the derivative of the resistance. We fit $R_{\text{total}}(V_{g1}, V_{g2})$ for $V_{g1,2} \geq -4$ V to more precisely account for the feature that will determine the manifold sign change of the thermoelectric current for $V_{g1,2} \geq 0$ V. Generally, a numerical evaluation of the derivative $\frac{\partial R_{\text{total}}}{\partial V_g}$ results in a noisy computed Seebeck coefficient. Figure S5a and S5c depict the numerical results of $\frac{\partial R_{\text{total}}}{\partial V_{g2}}$ and $\frac{\partial R_{\text{total}}}{\partial V_{g1}}$ based on the experimental $R_{\text{total}}$, respectively. For comparison, Figures S5b and S5d depicts $\frac{\partial R_{\text{total}}}{\partial V_{g2}}$ and $\frac{\partial R_{\text{total}}}{\partial V_{g1}}$ that are calculated from the fit of $R_{\text{total}}(V_{g1}, V_{g2})$. The charge neutrality point is identical for both methods. Therefore, $R_{\text{total}}(V_{g1}, V_{g2})$ is fitted, and we use the resulting parameters to model the thermoelectric current according to equation eq.(3) of the main manuscript.